\documentclass[universe,article,submit,pdftex,moreauthors]{Definitions/mdpi}

\firstpage{1} 
\pubvolume{1}
\issuenum{1}
\articlenumber{0}
\pubyear{2023}
\copyrightyear{2023}
\datereceived{ } 
\daterevised{ } 
\dateaccepted{ } 
\datepublished{ } 
\hreflink{https://doi.org/} 

\Title{Using Neural Networks to Perform Rapid High-Dimensional Kilonova Parameter Inference}

\TitleCitation{Using Neural Networks to Perform Rapid High-Dimensional Kilonova Parameter Inference}

\Author{Mouza Almualla $^{1, \dagger,*}$\orcidA{}, Yuhong Ning $^{2, \ddagger}$, Pouyan Salehi$^{3}$, Mattia Bulla$^{4,5,6,7}$, Tim Dietrich$^{3,8}$, Michael W. Coughlin$^{9}$ and Nidhal Guessoum $^{1}$}

\AuthorNames{Mouza Almualla, Yuhong Ning, Pouyan Salehi, Mattia Bulla, Tim Dietrich, Michael Coughlin and Nidhal Guessoum}

\AuthorCitation{Almualla, M.; Ning, Y.; Salehi, P.; Bulla, M.; Dietrich, T; Coughlin, M; Guessoum, N}

\address{%
$^{1}$ \quad American University of Sharjah, Physics Department, PO Box 26666, Sharjah, UAE\\
$^{2}$ \quad Zhiyuan College/School of Mathematical Sciences, Shanghai Jiao Tong University, Shanghai 200240, China\\
$^{3}$ \quad Institute of Physics and Astronomy, University of Potsdam, Karl-Liebknecht-Str. 24/25, 14476, Potsdam, Germany\\
$^{4}$ \quad Department of Physics and Earth Science, University of Ferrara, via Saragat 1, I-44122 Ferrara, Italy\\
$^{5}$ \quad INFN, Sezione di Ferrara, via Saragat 1, I-44122 Ferrara, Italy\\
$^{6}$ \quad INAF, Osservatorio Astronomico d’Abruzzo, via Mentore Maggini snc, 64100 Teramo, Italy\\
$^{7}$ \quad The Oskar Klein Centre, Department of Astronomy, Stockholm University, AlbaNova, SE-106 91 Stockholm, Sweden\\
$^{8}$ \quad Max Planck Institute for Gravitational Physics (Albert Einstein Institute), Am M\"uhlenberg 1, D-14476 Potsdam, Germany\\
$^{9}$ \quad School of Physics and Astronomy, University of Minnesota, Minneapolis, Minnesota 55455, USA
}

\corres{Correspondence: mouza.almualla@cfa.harvard.edu}

\firstnote{Current address: Center for Astrophysics $\vert$ Harvard \& Smithsonian, 60 Garden St, Cambridge, MA 02138, USA} 
\secondnote{Current address: Department of Physics, Brown University, Providence, RI 02906, USA}

\abstract{On the 17th of August, 2017 came the simultaneous detections of GW170817, a gravitational wave that originated from the coalescence of two neutron stars, along with the gamma-ray burst GRB170817A, and the kilonova counterpart AT2017gfo. Since then, there has been much excitement surrounding the study of neutron star mergers, both observationally, using a variety of tools, and theoretically, with the development of complex models describing the gravitational-wave and electromagnetic signals.
In this work, we improve upon our pipeline to infer kilonova properties from observed light-curves by employing a Neural-Network framework that reduces execution time and handles much larger simulation sets than previously possible. In particular, we use the radiative transfer code \texttt{POSSIS} to construct 5-dimensional kilonova grids where we employ different functional forms for the angular dependency for the dynamical ejecta component. We find that incorporating an angular dependence improves the fit to the AT2017gfo light-curves by up to $\sim$50\% when quantified in terms of the weighted Mean Square Error.}


\begin{document}
\nolinenumbers


\section{Introduction}
\label{sec:intro}
The driving idea of multi-messenger astronomy is that a single event observed through different messengers improves our understanding of the physical processes. This concept has gotten additional momentum through the Binary Neutron Star (BNS) detection GW170817~\citep{AbEA2017b} connected with the observation of electromagnetic radiation from gamma-rays to radio waves \citep{AbEA2017h}. This event followed the birth of GW astronomy in 2015 \citep{AbEA2016}, brought about with the success of the Advanced Virgo \citep{adVirgo} and the Advanced LIGO \citep{aLIGO} interferometers in detecting GWs emitted by compact object mergers. In general, BNS and Neutron Star -- Black Hole (NSBH) mergers may be accompanied by an optical/infrared counterpart, referred to as a \textit{kilonova}, which will give us important insights into what is possibly one of the most dominant sites of rapid neutron capture (r-process) elements in the Universe; see~\cite{Me2017} for a review. 
Our first and as yet only confirmed discovery of a kilonova from a BNS merger, AT2017gfo in NGC 4993 (D $\sim40$ Mpc) \citep{2017Sci...358.1556C}, was achieved through follow-up of GW170817 by multiple telescope facilities around the globe, and is an exceptional demonstration of the potential of multi-messenger astronomy in contributing to scientific knowledge~\citep{AbEA2017h}. These detections were also accompanied by the discovery of short gamma-ray burst GRB 170817A \citep{AbEA2017e}. Through this event, we learned more about the aforementioned r-process nucleosynthesis \citep[e.g.,][]{ChBe2017,2017Sci...358.1556C, CoBe2017,PiDa2017,SmCh2017,WaHa2019,KaKa2019} and also placed constraints on the expansion rate of the Universe \citep{2017Natur.551...85A,Guidorzi:2017ogy,HoNa2018,CoDi2019,Dhawan:2020a,DiCo2020} and the neutron star equation of state \citep[e.g.,][]{BaJu2017, MaMe2017, CoDi2018b, CoDi2019b, AnEe2018, MoWe2018,RaPe2018,AbAb2018,Lai2019,DiCo2020,HuPa2021}. 

Inferring kilonova model parameters from observational data requires a thorough understanding of the ejecta composition, behaviour, and morphology \citep[e.g.,][]{DiCo2020,Nedora:2020qtd,NiMa2021,BrPe21}; in addition, there are various parts of the parameter inference pipeline that must be improved upon and optimized such that we are able to efficiently utilize the available data and learn more about the progenitors. Radiative transfer simulations have been most commonly used to generate expected supernova and kilonova light-curves, \citep[e.g.,][]{Tanaka2016,KaMe2017,Wollaeger2018,Bu2019,kawa2020}.

There are two major sources of ejecta from the progenitor system. The first is known as the \textit{dynamical ejecta}, ejected `dynamically' around the moment of merger through torque or shocks (at merger or in the early-postmerger phase) \citep[e.g.,][]{MePi2008,BaGo2013}.
The second source of ejecta comes from the \textit{post-merger wind}, i.e., outflows from an accretion disk formed around the central remnant, consisting of the unbound debris resulting from the merger \citep[e.g.,][]{SiMe2018,Metzger2019}. The dynamical source is ejected on a timescale of milliseconds, while the latter is formed on a timescale of up to seconds.


There are two major dynamical ejection processes in BNS mergers. The first is through shocks that forms at the contact interface of the merging stars or at core bounces in the early postmerger phase~\citep{Hotokezaka_2015}. The second component is the tidal ejecta, and is due to tidal interactions that arise from the non-axisymmetric gravitational forces at play in the binary system \citep{Hotokezaka_2015,Metzger2019}. Recent simulations~\citep{Radice:2018pdn,Nedora:2020qtd} show that the more neutron-rich, lower $Y_{e}$ ejecta ($Y_{e} \lesssim 0.1$) arises from the aforementioned tidal component, and is located closer to the equatorial plane, while the ejecta arising from the shock component tends to have a relatively higher electron fraction (going up to $Y_{e}\sim\,0.4$) and is approximately isotropic. 

The disk mass is $\sim 0.01-0.3 M_{\odot}$ \citep{Oechslin_2006}, and is usually much lower when the remnant immediately collapses into a black hole; this is due to the lack of time for the remnant to redistribute its angular momentum and mass as it transforms from a differentially rotating to a solid rotating body, thus preventing the formation of a more massive disk \citep{Metzger2019}. A fraction of the disk is ejected in the form of a post-merger disk wind, with the exact value being uncertain and ranging from 20 to 40~per cent \citep[e.g.][]{JuBa2015,SiMe2018,Miller2019,Fernandez2019,Fujibayashi2020}. The post-merger wind usually dominates the dynamical ejecta \citep{WuFe2016}. Works such as \cite{FeMe2013} show the disk wind ejecta $Y_{e}$ ranging within $\sim 0.2-0.4$ when the remnant collapses into a black hole. The electron fractions present in the case of both dynamical and post merger wind cases are suitable for the production of heavy r-process elements~\citep{Rosswog:2013kqa}.

With the gravitational-wave detector networks' fourth observing run nearing, one needs to continue to improve the instruments and software tools that we use to analyze the forthcoming multi-messenger detections. These mergers bring rich knowledge to various fields of physics and astronomy, from the understanding of such events and the big, powerful jets of relativistic particles that they produce, to the physical structure of both the neutron stars (internally) and the matter that is ejected in the merger, not to mention the detailed characteristics of the gravitational waves that are emitted. Much work is being done on the theoretical front to prepare us for the `treasure trove' of data from upcoming detections. In particular, models of kilonovae are being constructed with increased complexity and fine-tuning \citep[e.g.,][]{Ko2021,Zhu2021,WoCha2021}.

In this work, we use \texttt{POSSIS} \citep{Bu2019}, a Monte-Carlo radiative transfer code, in order to produce ``model grids'' that sample the parameter space for any given ejecta morphology. In particular, we use an updated version of \texttt{POSSIS} \citep{possis2} including improved heating rates from \cite{Rosswog2022} and wavelength- and time-dependent opacities from \cite{Tanaka2020} that depend on local properties of the ejecta (density, temperature and $Y_e$). Details about these new implementations can be found in \cite{possis2}. Previous works such as \cite{CoDi2018,DiCo2020,Heinzel2021} used Gaussian Process Regression (GPR) to interpolate within kilonova grids to perform parameter estimation. Here, we introduce a Neural Network (NN) framework that significantly enhances the speed of the interpolation step and scales well for much larger grids. In addition, due to the importance of understanding the effects of different ejecta morphologies on the computed light-curves, we explore different functional forms for the angular dependence in the density profile of the dynamical ejecta. 

Our paper is structured as follows, in Sec.~\ref{sec:nnarch}, we describe the NN architecture. In Sec.~\ref{sec:angdep}, we discuss and evaluate the different density profiles for the dynamical ejecta in the context of GW170817/AT2017gfo. Finally, we summarize our conclusions and outline ways of moving forward in Sec.~\ref{sec:conclusion}.
\section{Methods: A New Surrogate Generation Framework}

\label{sec:nnarch}
\begin{figure}[h]
    \centering
    \includegraphics[width=\columnwidth]{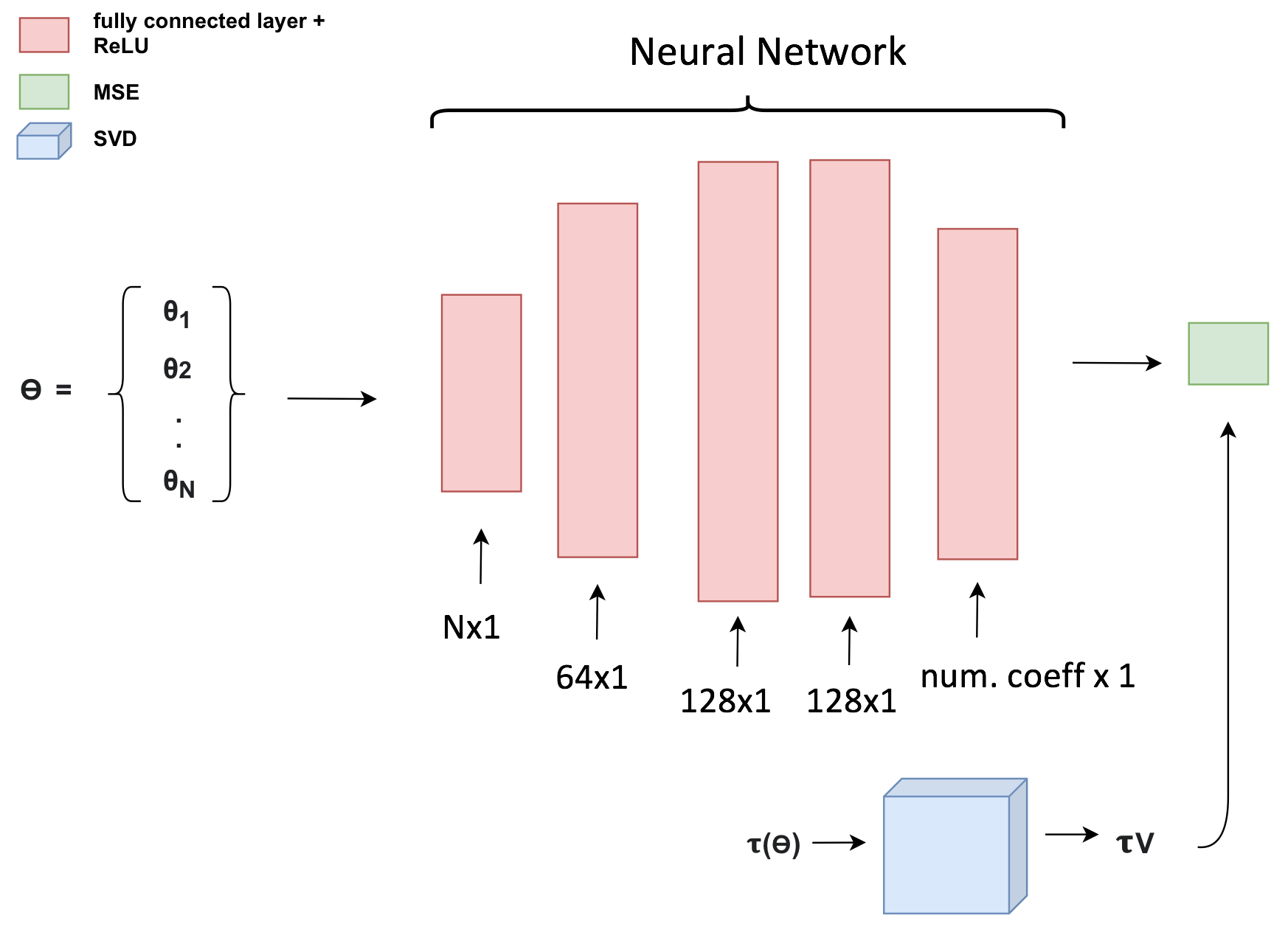}
    \caption{A diagram of the interpolation framework used in this work. The input to our network is the model parameters, and the network is trained on a PCA-like projection of our photometric and bolometric data onto the right-singular vector. Five fully-connected layers are shown, and an MSE loss function is employed.}
    \label{fig:interpfw}
\end{figure}
\begin{figure*}[bhpt]
    \centering
    \includegraphics[width=\columnwidth]{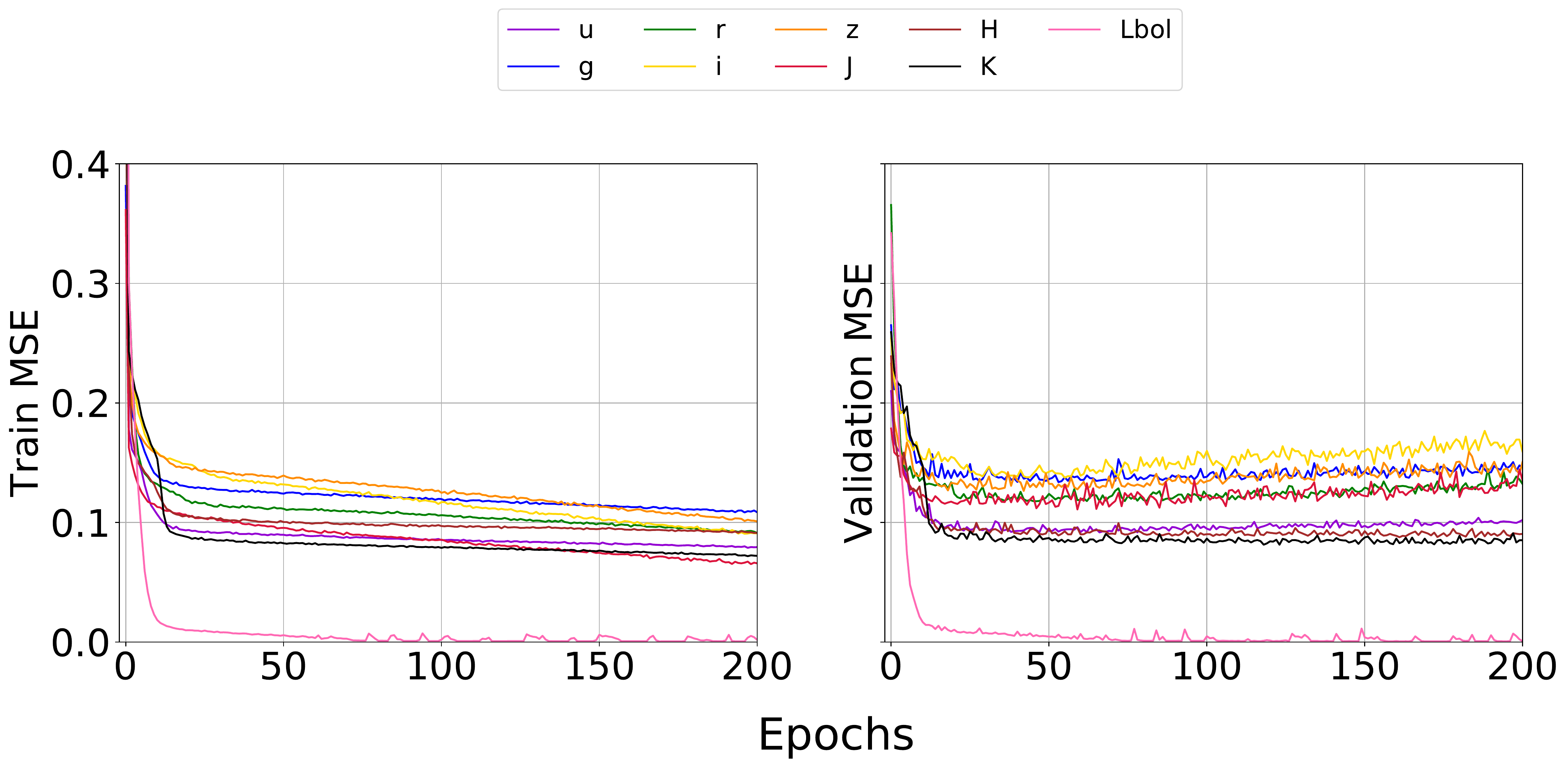}
    \caption{The MSE loss is shown for the $u$, $g$, $r$, $i$, $z$, $y$, $J$, $H$, and $K$ bands, as well as for the bolometric luminosity. The loss for the train set is shown in the left subplot, and for the validation set in the right subplot. After $\sim$ 200 epochs, the network tends to over-fit the train data, so we stop training at this point.}
    \label{fig:MSEloss}
\end{figure*}
\begin{figure}
    \centering
    \includegraphics[width=\columnwidth]{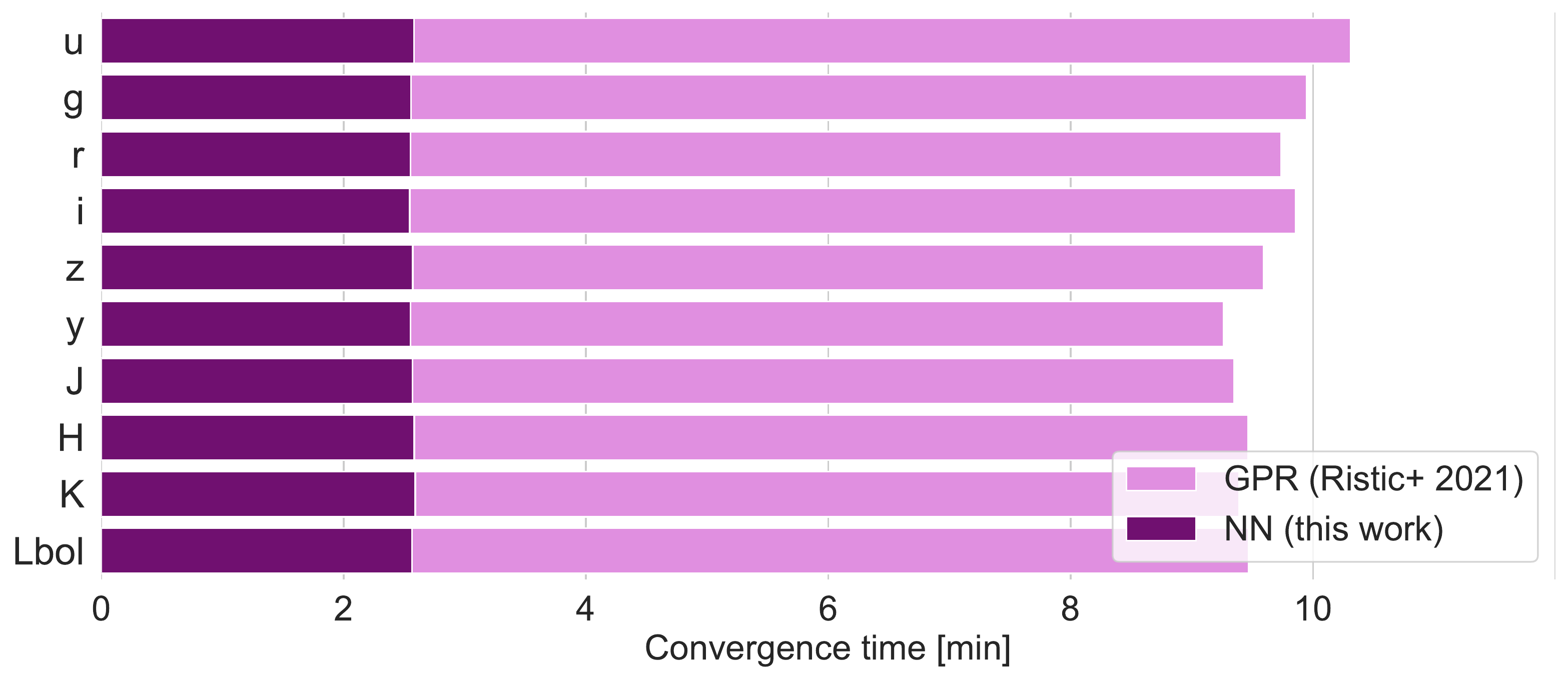}
    \caption{Convergence time for all of the photometric bands used in this work, as well as for the bolometric luminosity. The NN framework is shown in dark purple, while the interpolation scheme from \cite{ristic2021interpolating} is shown in light purple.}
    \label{fig:exec_time}
\end{figure}
\subsection{Preprocessing}
Instead of directly inputting our bolometric and photometric light-curves into our Neural Network for interpolation, we follow the procedure of~\cite{DoFa2017} and~\cite{CoDi2018}, using Singular Value Decomposition (SVD) to perform a dimensionality reduction on our data and produce the principle components of our light-curve vectors, i.e., a new basis for our data in which different dimensions of the data are most uncorrelated. We then train our neural network on this output in order to produce a so-called ``surrogate model'' that interpolates our original model grid. Denoting our photometric and bolometric light-curves as $\tau(\mathbf{\Theta})$, where $\mathbf{\Theta}$ represents the set of parameters for a given model, SVD decomposes $\tau(\mathbf{\Theta})$ as follows:
\begin{equation}
    \tau = U\Sigma V^{T}
\end{equation}
where $\Sigma$ is a diagonal matrix containing the singular values of $\tau(\mathbf{\Theta})$, and the columns of U and V are orthonormal bases called the left-singular and right-singular vectors, respectively. We can then obtain the principle component analysis (PCA) output by projecting $\tau(\mathbf{\Theta})$ onto the right singular matrix:
\begin{equation}
     \tau V = PCA
\end{equation}
Note that we first normalized our light curves before performing PCA, and the right singular vector $V$ was truncated to 100 basis vectors. A train/validation split of 90/10 was used.
\subsection{Neural Network Architecture}

Our NN consists of five ReLU-activated (Rectified Linear Units, i.e., $R(z) = \text{max}(0,z)$ for each neuron) fully-connected layers, as shown in Figure~\ref{fig:interpfw}, with the first layer containing $N$ weights corresponding to the number of parameters for the model, three hidden layers, and the output layer having a number of weights equal to the number of columns in our PCA output. We used a batch size of 32 and trained for 200 epochs, using a mean-square error (MSE) loss function.

The use of a NN is much more efficient for higher-dimensional kilonova models than GPR, which usually performs poorly when a model has more than four parameters. To demonstrate the performance of our interpolation framework, we use a model (to be described in-depth in the following sections) that has 5 parameters and 1,260 total parameter combinations comprising our simulation set. Training 10 different networks (one for each photometric band $u$, $g$, $r$, $i$, $z$, $y$, $J$, $H$, $K$, and in addition one for the bolometric luminosity) took on the order of $\sim$ 40 mins. The MSE loss is shown as a function of epoch in Figure~\ref{fig:MSEloss}. Our model does not over- or under-fit the train data, since we can see very similar performance on our validation set in the right subplot as compared to the train set. In addition, our network converges very quickly, so even 50 epochs would be sufficient in constructing our surrogate. 

We also compare our NN framework to state-of-the-art interpolation methods, such as the GPR framework used in \cite{CoDi2018,Heinzel2021}, as well as that in \cite{ristic2021interpolating}, in terms of its computational efficiency. It took upwards of 13 hours for the former GPR implementation from \cite{CoDi2018} (which uses the \texttt{scikit-learn} library) to construct our model in the $u$-band, and so we only use the framework from \cite{ristic2021interpolating} in this comparison. The performance for each model is shown in Figure~\ref{fig:exec_time}. In addition to the benefits in terms of convergence time, the output of the NN interpolation scheme takes up a lot less memory -- on the order of tens of MB, in comparison to a few GBs for the \texttt{scikit-learn} GPR scheme. Even more importantly, the total time it takes to perform parameter inference is significantly reduced due to the much simpler loading procedure for NNs; for the 5-dimensional model used here, loading for all of the bands and the bolometric luminosity takes a total of $\sim$700\,ms for the NN, while for the GPR approach from \cite{ristic2021interpolating}, it takes $\sim$\,80\,min (since some components need to be recalculated\footnote{There may be some room for improvement in loading for the GPR method, so consider this an upper limit on the execution time.}). This is a significant improvement when considering that the parameter inference takes $\approx$20\,min excluding the loading step.

Future work in improving the performance of our model includes incorporating batch normalization, drop out layers, and various other easily implementable techniques. The wealth of developments in deep learning allows for a significant improvement in computational efficiency whilst still maintaining high accuracies and developing surrogate models with high fidelity.

\section{Results: Angular Dependencies in the Dynamical Ejecta}
\label{sec:angdep}

Numerical relativity simulations provide valuable information about what we can expect from the ejected matter in the merger of BNSs. As the field develops, these simulations have become more and more complex, incorporating neutrino transport and more realistic neutron star equations of state. We know that there are two primary ejection processes, described in Section~\ref{sec:intro}, but the exact geometry of each of these components are now known; note that also the ejecta opacities are another point of uncertainty in kilonova simulations and we refer to Appendix~\ref{appendix:scale} for a preliminary exploration of this. Generally, the density profile of the post-merger wind is thought to be relatively spherically symmetric, while the density profile of the dynamical ejecta will have an angular dependence such that the mass is more concentrated in equatorial regions as compared to polar regions \citep[e.g.,][]{dietrich2017,kawa2020,Nedora:2020qtd}.

There are various functional forms that have been used in kilonova modelling to incorporate this angular dependency into the dynamical ejecta. We will explore a few of them, and then use data from AT2017gfo to infer which provides the best fit to the observed light-curves. We remark that in our simulations, we will only consider the case in which the post-merger wind is slower than the dynamical ejecta. We use the so-called \textit{Spherical segment-Spherical Cap} geometry (shortened as SSCr, with $r$ signifying that re-processing between the different ejecta components is taken into account); see the left-most geometry in Figure 1 of \cite{Heinzel2021}. The density profiles for the different ejecta components are represented as follows:

\[ \rho(r,t) \propto \begin{cases} 
      r^{-\alpha}t^{-3} & v_{\text{pm,min}}\leq r/t \leq v_{\text{pm,max}} \\
      \eta(\theta)r^{-\alpha}t^{-3} & v_{\text{pm,max}}\leq r/t \leq v_{\text{dyn,max}} 
   \end{cases}
\]
where we set $\alpha$=3, $v_{\text{pm,min}}=0.025c$, $v_{\text{pm,max}}=0.08c$, $v_{\text{dyn,max}}=0.3c$. $\eta(\theta)$ is the term introducing the angular dependence with respect to $\theta$, the angle measured from the polar axis, into the dynamical ejecta. The forms that we will first explore are: 

1) a sinusoidal relationship, i.e., $\eta(\theta)=\sin^{2}(\theta)$, also used by \cite{PeRa2017} based on numerical-relativity simulations from \cite{Ra2018}; and 

\begin{figure}[htb]
    \centering
    \includegraphics[width=\columnwidth]{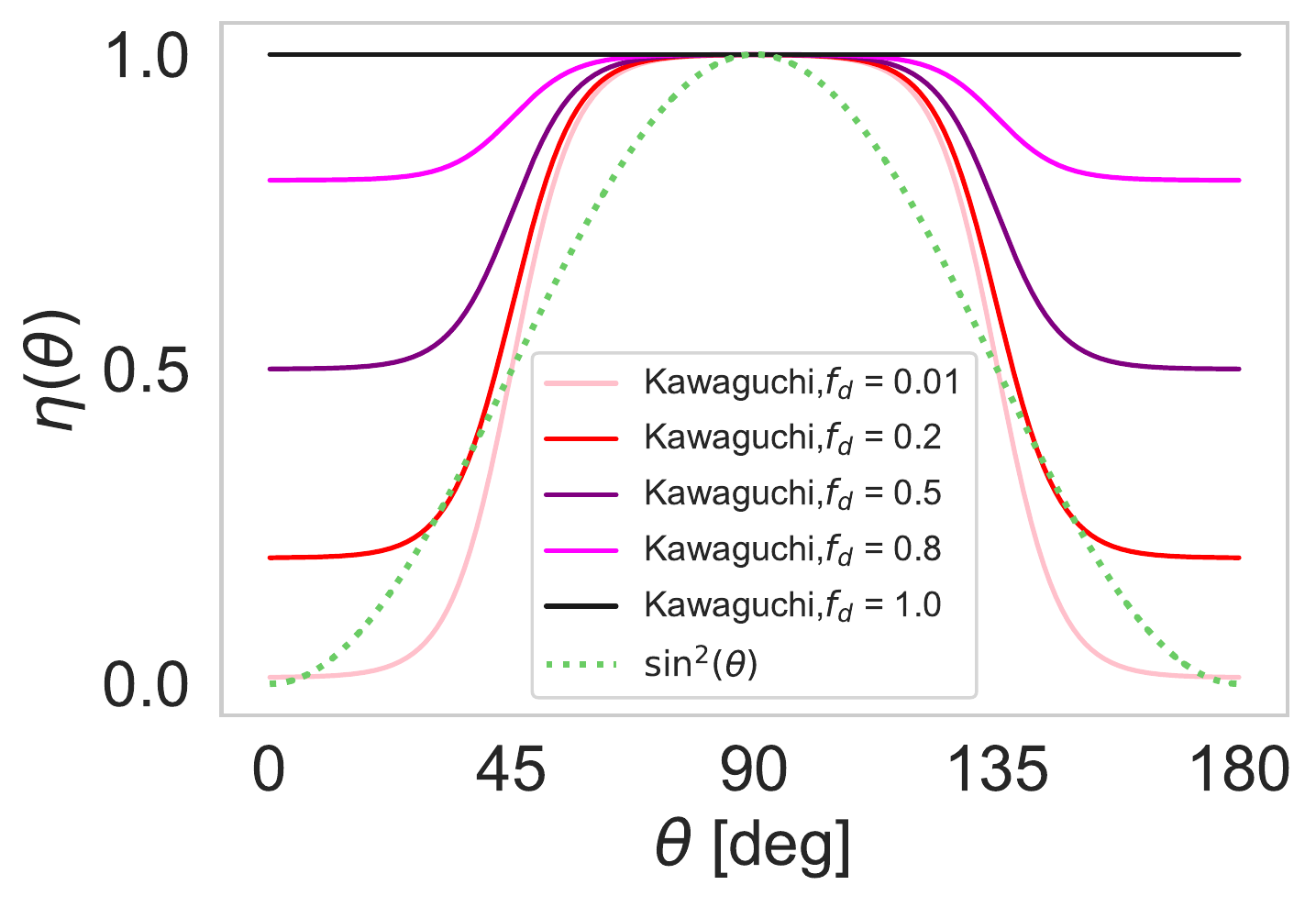}
    \caption{Different functional forms for the angular dependency in the dynamical ejecta. This is shown for $\sin^{2}(\theta)$ and Equations~\eqref{ref:eq_eta} and~\eqref{ref:eq_theta}. $\theta$ represents the angle from the polar axis. }
    \label{fig:angulardep}
\end{figure}

2) a function defined in \cite{kawa2020}:
\begin{equation}
    \eta(\theta) = (1-\Theta(\theta))f_{\text{d}}+\Theta(\theta) \label{ref:eq_eta}
\end{equation}
\begin{equation}
    \Theta(\theta) = \frac{1}{1+\exp{[-10(\theta-\pi/4])]}} \label{ref:eq_theta}
\end{equation}
where $f_{d}$ determines the strength of the angular dependence from equatorial to polar angles. We will call the latter the ``Kawaguchi Model.'' As can be seen in Figure~\ref{fig:angulardep}, with the sinusoidal functions, we have a very gradual increase, whereas with the dependency from \cite{kawa2020} the density profile will be much more concentrated for $\pi/4 \leq \theta \leq 3\pi/4$. 
\begin{figure*}[htbp]
    \centering
    \includegraphics[width=\textwidth]{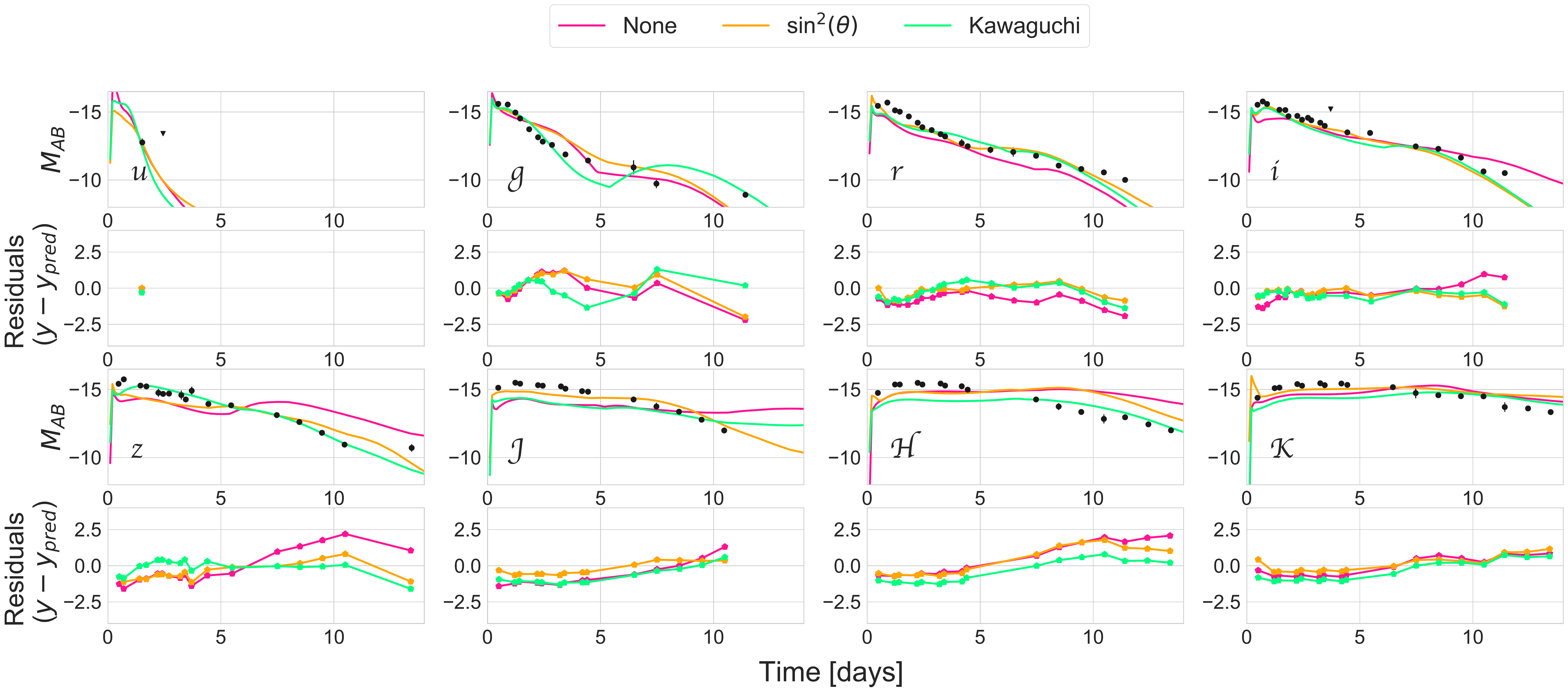}
    \caption{Best-fit light-curves in the $u$, $g$, $r$, $i$, $z$, $J$, $H$, and $K$ bands are plotted for three different models in which the angular dependency in the dynamical ejecta is varied. The AT2017gfo observations are plotted as black dots, the Kawaguchi best-fit is plotted in green, the $\sin^{2}(\theta)$ best-fit is plotted in orange, and the model with no angular dependency is plotted in pink. We also plot the residuals in each band.}
    \label{fig:basicangdep}
\end{figure*}
For our first analysis here, we will be pinning $f_{d}$ at 0.01 as done in~\cite{kawa2020}, since this produces the strongest angular dependency and will be a useful comparison. Our model has the following four free parameters: the dynamical ejecta mass $M_{\text{ej,dyn}}$, post-merger wind ejecta mass $M_{\text{ej,dyn}}$, half-opening angle of the lanthanide-rich component of the dynamical ejecta $\phi$, and the observing angle $\theta_{\text{obs}}$. For the dynamical and post-merger wind ejecta masses, we perform logarithmic sampling across the interval [0.001, 0.1]. Our entire sample set used to generate our model grids is shown in Table~\ref{tab:table1}.

\begin{table}[H]
\caption{The parameters of our model and the sample values for which we run \texttt{POSSIS}.}
\label{tab:table1}
\newcolumntype{C}{>{\centering\arraybackslash}X}
\begin{tabularx}{\textwidth}{C|C}
\textbf{Parameters} & {\textbf{Samples}} \\
\toprule 
$M_{\text{ej,dyn}}$/$M_{\odot}$ & [0.001, 0.00251, 0.00631, 0.0158, 0.0398, 0.1]\\
$M_{\text{ej,win}}$/$M_{\odot}$ &  [0.001, 0.00251, 0.00631, 0.0158, 0.0398, 0.1] \\
$\phi$ [deg] & [0, 15, 30, 45, 60, 75, 90]\\
$\cos{(\theta_{\text{obs}})}$ & [0, 0.1, 0.2, 0.3, 0.4, 0.5, 0.6, 0.7, 0.8, 0.9, 1.0] 
\end{tabularx}
\end{table}

We generate 3 model grids: 1) no angular dependence, 2) angular dependence $\eta=\sin^{2}(\theta)$, and 3) the Kawaguchi model with $f_{d}$=0.01. We thus run \texttt{POSSIS} over the aforementioned parameter space, generating $6\times6\times7=252$ simulations for each model, and then used the NN interpolation described in Section~\ref{sec:nnarch} to interpolate the grid. We then perform parameter inference and find the maximum likelihood estimates for AT2017gfo using \texttt{PyMultiNest}, which is capable of performing both model selection and parameter inference. Using \texttt{PyMultiNest}, we extract posterior distributions for each of the aforementioned parameters. We assume uniform priors that extend over the limits of the grid in the parameter space. The results of the inference (i.e., the best-fit light-curves), assuming $\pm 1$\,mag systematic error, for each of these models are shown in Figure~\ref{fig:basicangdep}.

Inspecting by eye, we can see in Figure~\ref{fig:basicangdep} that across all bands the $\sin^{2}(\theta)$ and Kawaguchi models result in residuals closer to 0. We can further quantify this by obtaining the MSE for each model. However, since the Local Thermodynamic Equilibrium (LTE) assumption in \texttt{POSSIS} is likely to fail at late times when the ejecta become optically thin, and the brighter early-time emission is more important for observational purposes, it is reasonable to perform a Weighted MSE (WMSE) prioritizing the early observations. We thus use the following equation:
\begin{equation}
    \text{WMSE} = \frac{1}{n} \frac{\sum_{i=0}^{n-1} \omega_{i}(M_{\text{obs}}-M_{\text{pred}})^{2}}{\sum_{i=0}^{n-1}\omega_{i}}
\end{equation}
where $\omega_{i} = \ln{(|M_{\text{obs}}|+1)}+1$. Having done so, we obtain the WMSE values shown in Table~\ref{tab:mse_1}, and verify our conclusion that incorporating some form of an angular dependence improves the fit to AT2017gfo across almost all bands.

\begin{figure}[!htbp]
    \centering
    \includegraphics[width=\columnwidth]{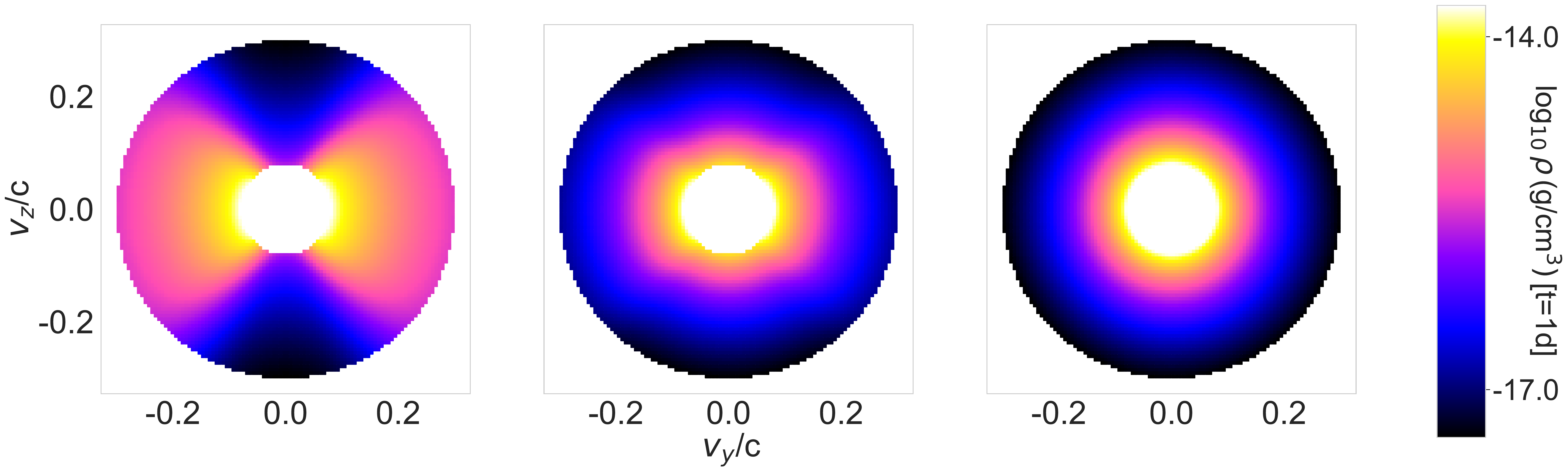}
    \caption{2D dynamical ejecta density profiles for $f_{d}$=0.01, $f_{d}$=0.5, $f_{d}$=1.0.}
    \label{fig:densplt_gnuplot2}
\end{figure}

The parameter $f_{d}$ allows us also to assess which dependence provides the best fit to the  measured light-curves. In Figure~\ref{fig:densplt_gnuplot2}, we show the 2D density profile for different $f_{d}$'s (0.01, 0.5, 1.0) for comparison.

\begin{figure}
    \centering
    \includegraphics[width=\columnwidth]{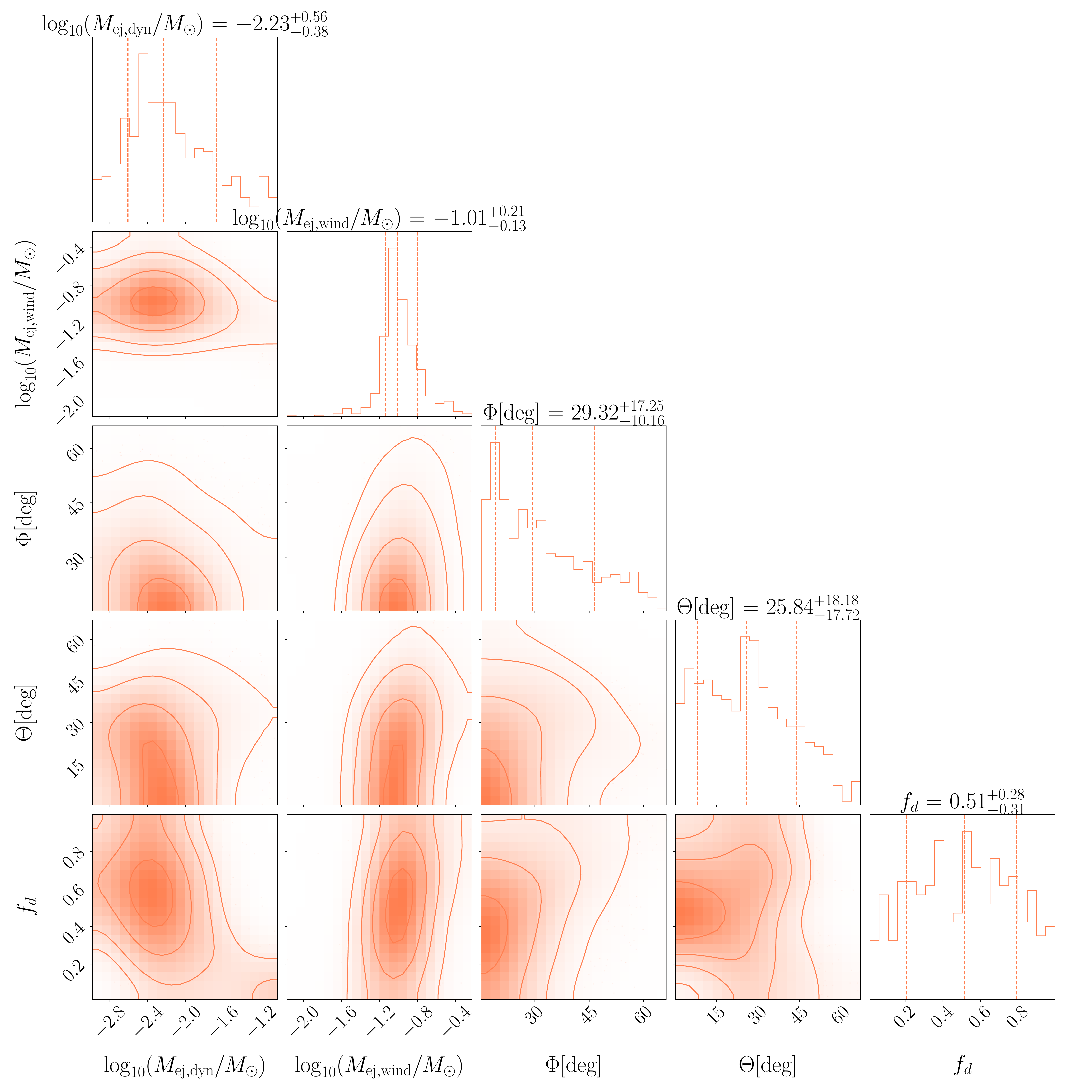}
    \caption{Corner plot showing the resulting posteriors for our model with $f_{d}$ as a free parameter. We can see that all of our inferred parameters are within reasonable ranges and that the distribution for $f_{d}$ is centered around 0.50.}
    \label{fig:corner_fd}
\end{figure}

For this analysis, we include $f_{d}$ as an additional parameter to those listed in Table~\ref{tab:table1}, sampling $f_{d}=[0.01, 0.2, 0.5, 0.8, 1.0]$. This increases our simulation set to 1260 simulations. The resulting corner plot of our posteriors is shown in Figure~\ref{fig:corner_fd}. We also include the WMSE values with our maximum-likelihood $f_{d}$ in Table~\ref{tab:mse_1}.

\begin{table}[H]
\caption{WMSE values for each band and model, for when no angular dependence is incoporated, versus a $\sin^{2}(\theta)$ dependency, versus the Kawaguchi equation when pinning $f_{d}$ at 0.01, and finally when allowing $f_{d}$ to vary (i.e., $f_{d}=0.50$).}
\label{tab:mse_1}
\newcolumntype{C}{>{\centering\arraybackslash}X}
\begin{tabularx}{\textwidth}{C|C|C|C|C}
\textbf{Band} & \multicolumn{4}{c}{\textbf{Models}} \\ 
\toprule
& None & $\sin^{2}(\theta)$ & $f_{d}=0.01$ & best-fit $f_{d}$ \\
\midrule
$g$	&	0.0642	&	0.0571	&	\textbf{0.0291}	&	0.0738	\\
$r$	&	0.0461	&	\textbf{0.0127}	&	0.0204	&	0.0227	\\
$i$	&	0.0242	&	0.0119	&	0.0154	&	\textbf{0.0034}	\\
$z$	&	0.0771	&	0.0313	&	0.0160	&	\textbf{0.0109}	\\
$J$	&	0.0781	&	\textbf{0.0164}	&	0.0649	&	0.0317	\\
$H$	&	0.0846	&	0.0593	&	0.0506	& \textbf{0.0412}	\\
$K$	&	0.0249	&	\textbf{0.0168}	&	0.0370	&	0.0340	\\
\textbf{average} & 0.0570 & 0.0294 & 0.0333 & 0.0311\\
\end{tabularx}
\end{table}

We can see that $f_{d}\sim\,0.50^{+0.28}_{-0.32}$ provides the most optimal fit. When fitting for $f_{d}$, the previous WMSE of 0.0333 decreases to 0.0311. The fit is significantly better in redder bands with $f_{d}$=0.50, while it would be better using a stronger dependency $f_{d}=0.01$ with respect to bluer $g$ and $r$ bands. In order to confirm this behavior, we also run the inference with an assumed systematic error of 0.5\,mag (instead of 1.0), and find that the distribution is even more clearly distributed around $f_{d}\sim0.6$, and the WMSE further decreases to 0.0252. A moderate angular dependence is thus preferred.

\begin{table}[H]
\caption{Resultant posteriors for $M_{\text{ej,pm}}$, $M_{\text{ej,dyn}}$, $\phi$, and $\theta_{\text{obs}}$ for each of the models explored in this section. Uncertainties are shown at the 15th and 85th percentiles.}
\label{tab:paramposteriors}
\newcolumntype{C}{>{\centering\arraybackslash}X}
\begin{tabularx}{\textwidth}{C|CCCC}
\textbf{Model} & \multicolumn{4}{c}{\textbf{Parameters}} \\
$\eta(\theta)$& $M_{\text{ej,dyn}}$ [$\times 10^{-3} M_{\odot}$]& $M_{\text{ej,pm}}$ [$M_{\odot}$]& $\phi$ [deg] & $\theta_{\text{obs}}$ [deg] \\
\toprule
$1$ & $4.57_{-2.48}^{+21.88}$ & $0.110_{-0.0302}^{+0.0488}$ & $44.29_{-13.69}^{+11.85}$  & $32.01_{-19.00}^{+19.38}$  \\
& & & & \\
$\sin^{2}(\theta)$ & $3.09_{-1.51}^{+18.29}$ & $0.0912_{-0.032}^{+0.209}$ & $36.03_{-12.77}^{+14.04}$ & $24.86_{-17.43}^{+24.89}$ \\
& & & & \\
kawaguchi ($f_{d}=0.01$) & $19.05_{-11.98}^{+35.90}$ & $0.0832_{-0.0405}^{+0.0456}$ & $26.95_{-8.03}^{+12.28}$ & $37.87_{-19.50}^{+15.77}$ \\
& & & & \\
kawaguchi (varied $f_{d}$) & $5.89_{-3.43}^{+15.49}$ & $0.0977_{-0.0236}^{+0.0645}$ & $29.32_{-10.16}^{+17.25}$ & $25.84_{-17.72}^{+18.18}$ \\
& & & & \\
\end{tabularx}
\end{table}


To finalize our exploration of different angular dependencies, we summarize the inferred constraints on $M_{\text{ej,pm}}$, $M_{\text{ej,dyn}}$, $\phi$, and $\theta_{\text{obs}}$ for each model in Table~\ref{tab:paramposteriors}.

\section{Conclusions}
\label{sec:conclusion}
In this work, we have modified and improved upon the present parameter inference pipeline used to infer ejecta parameters from observed kilonova light-curves \citep{CoDi2018,DiCo2020}. We constructed an NN framework that sufficiently models the sampled lightcurves and produces a surrogate model with a much improved computational efficiency as compared to previous surrogate construction methods.

We explored the possible nature of AT2071gfo's ejecta geometry. Specifically, we looked at the angular dependence in the density profile of the dynamical ejecta, showing that some form of angular dependence is beneficial and provides better fits to GW710817/AT2017gfo, with the $\sin^{2}(\theta)$ and the Kawaguchi functional forms both outperforming the model with no angular dependence. The $\sin^{2}(\theta)$ dependence improves the fit across all bands by $\sim$50\%.

When using the Kawaguchi equation -- parameterized by $f_{d}$, which is approximately the ratio of the density in the polar region ($0\leq\theta\leq \pi/4$) to that in the equatorial region ($\pi/4\leq\theta\leq \pi/2$) -- to study the angular dependence, we found that $f_{d}\sim0.50$ is most suitable. Fitting for $f_{d}$ especially improves performance in the redder bands, decreasing the WMSE by 34\% when considering lightcurves in the $i$-band and onwards. 

Although we explored one specific aspect of the ejecta structure in this work, and there have been other recent explorations of different kilonova ejecta morphologies \citep{DaKa2020,Heinzel2021,Ko2021}, there is still much to explore. There are countless other geometries that could be considered for the kilonova ejecta, so an interesting path forward would be performing inference on a ``super-geometry'' that samples a much wider range of possible ejecta morphologies. 

\vspace{6pt} 



\authorcontributions{Conceptualization, M.A., Y.N., M.B., T.D.; methodology, M.A., Y.N.; software, M.B., Y.N., M.A.; validation, Y.N., M.B.; formal analysis, M.A., T.D., M.C., N.G.; investigation, M.A.; resources, M.B.; data curation, M.A.; writing---original draft preparation, M.A.; writing---review and editing, Y.N., M.B., T.D., M.C., N.G.; visualization, M.A.; supervision, M.B., T.D., M.C., N.G.; project administration, M.B., T.D., M.C., N.G.; funding acquisition, M.B., T.D., M.C. All authors have read and agreed to the published version of the manuscript.}

\funding{M.B.~acknowledges support from the Swedish Research Council (Reg. no. 2020-03330) and from the European Union’s Horizon 2020 Programme under the AHEAD2020 project (grant agreement n. 871158). M.W.C.\ acknowledges support from the National Science Foundation with grant Nos. PHY-2010970 and OAC-2117997. T.D.\ acknowledges financial support through the Max Planck Society.}

\dataavailability{The data presented in this study are openly available at \hreflink{https://github.com/mbulla/kilonova_models.}}

\acknowledgments{ We thank Stephan Rosswog, Oleg Korobkin, and Masaomi Tanaka for sharing the heating rate libraries and opacities used in this work. We are grateful for computational resources provided by the Leonard E Parker
Center for Gravitation, Cosmology and Astrophysics at the University of
Wisconsin-Milwaukee, as well as by the Supercomputing Laboratory at King Abdullah University of Science and Technology (KAUST) in Thuwal, Saudi Arabia.}

\conflictsofinterest{The authors declare no conflict of interest. The funders had no role in the design of the study; in the collection, analyses, or interpretation of data; in the writing of the manuscript; or in the decision to publish the results.} 





\appendixtitles{yes} 
\appendixstart
\appendix
\section[\appendixname~\thesection]{Introducing an Opacity-Scaling Parameter}
\label{appendix:scale}
Our simulations use state-of-the-art opacities for r-process elements \citep{Tanaka2020}. However, uncertainties in the opacities are likely to be present due to the challenges in modelling the myriad of line transitions expected for these heavy elements. 

Given these inherent uncertainties, we introduce the free parameter $\alpha_{\kappa}$ that scales the opacities, and in turn, fine-tunes the emitted flux and color of the kilonova. We assume, for simplicity, that the relative uncertainties are constant across time and wavelength, although this is likely to be more complicated in reality.
\begin{figure}[htbp]
    \centering
    \includegraphics[width=\columnwidth]{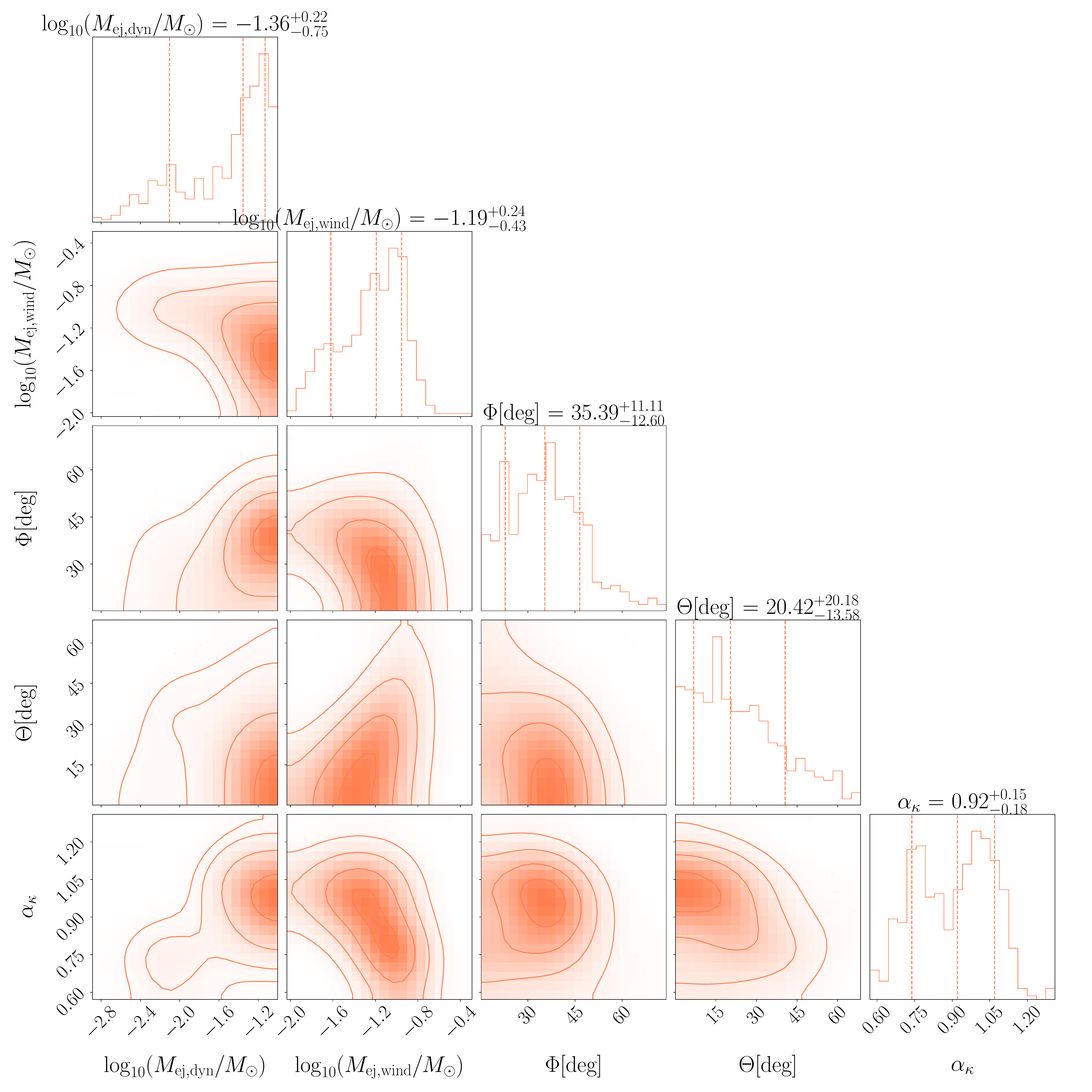}
    \caption{Posteriors showing that $\alpha_{\kappa}$ has a bimodal distribution when fit to GW170817/AT2017gfo lightcurves and a $\sin^{2}(\theta)$ dependency is assumed.}
    \label{fig:corner_kappa}
\end{figure}
In exploring this, we use the $\sin^{2}(\theta)$ model, given its good performance in Section~\ref{sec:angdep}. We use the same sample set as in Table~\ref{tab:table1}, including our new parameter $\alpha_{\kappa}$=[0.75,1.0,1.25]. The resulting corner plot is shown in Figure~\ref{fig:corner_kappa}, and we can clearly see a bimodal distribution, with one peak centered at $\alpha_{\kappa}\sim$\,1.04, and one centered at $\sim$\,0.75.




Hence, while at first glance it seems that introducing an opacity rescaling might improve the performance of the model, more work is required for a reliable estimation of the opacities. In particular, the strong bimodal nature of the distribution indicates that a simple time-independent rescaling of the opacities, as employed in this work, might not be applicable. And so, further work exploring either the temporal or the wavelength dependence of the ejecta opacities would be interesting, and may serve to limit the associated systematic errors.


\begin{adjustwidth}{-\extralength}{0cm}

\reftitle{References}




\bibliography{references.bib}

\PublishersNote{}
\end{adjustwidth}
\end{document}